\documentclass[conference]{IEEEtran}
\IEEEoverridecommandlockouts
\usepackage{cite}
\usepackage{amsmath,amssymb,amsfonts}
\usepackage{algorithmic}
\usepackage{graphicx}
\usepackage{textcomp}
\usepackage{xcolor}
\usepackage{booktabs}
\usepackage{multirow}
\usepackage{xspace}
\usepackage{caption}
\usepackage{subcaption}
\usepackage{footnote}
\usepackage{url}
\usepackage{tcolorbox}
\usepackage{flushend}

\def\BibTeX{{\rm B\kern-.05em{\sc i\kern-.025em b}\kern-.08em
    T\kern-.1667em\lower.7ex\hbox{E}\kern-.125emX}}

\begin{document}

\title{Automated Identification of\\On-hold Self-admitted Technical Debt}

\author{
\IEEEauthorblockN{Rungroj Maipradit\IEEEauthorrefmark{1},
Bin Lin\IEEEauthorrefmark{2},
Csaba Nagy\IEEEauthorrefmark{2}\\
Gabriele Bavota\IEEEauthorrefmark{2},
Michele Lanza\IEEEauthorrefmark{2}, 
Hideaki Hata\IEEEauthorrefmark{1},
Kenichi Matsumoto\IEEEauthorrefmark{1}}
\IEEEauthorblockA{
	\IEEEauthorrefmark{1}\emph{Nara Institute of Science and Technology, Japan}\\
	\IEEEauthorrefmark{2}\emph{Software Institute, USI Universit\`a della Svizzera Italiana, Switzerland}\\ 			
}}

\maketitle

\newcommand{\ie}{\emph{i.e.,}\xspace}
\newcommand{\eg}{\emph{e.g.,}\xspace}
\newcommand{\etc}{etc.\xspace}
\newcommand{\etal}{\emph{et~al.}\xspace}
\newcommand{\secref}[1]{Section~\ref{#1}\xspace}
\newcommand{\chapref}[1]{Chapter~\ref{#1}\xspace}
\newcommand{\appref}[1]{Appendix~\ref{#1}\xspace}
\newcommand{\figref}[1]{Fig.~\ref{#1}\xspace}
\newcommand{\listref}[1]{Listing~\ref{#1}\xspace}
\newcommand{\tabref}[1]{Table~\ref{#1}\xspace}
\newcommand{\tool}[1]{{\sc #1}\xspace}
\newcommand{\summary}[1]{{\begin{tcolorbox} \textbf{Summary}: #1 \end{tcolorbox}}}
\newcommand{\edit}[1]{{\textcolor{black}{#1}}}

\newboolean{showcomments}

\setboolean{showcomments}{true}

\ifthenelse{\boolean{showcomments}}
  {\newcommand{\nb}[2]{
    \fbox{\bfseries\sffamily\scriptsize#1}
    {\sf\small$\blacktriangleright$\textit{#2}$\blacktriangleleft$}
   }
  }
  {\newcommand{\nb}[2]{}
  }

\newcommand\GABRIELE[1]{\textcolor{blue}{\nb{GABRIELE}{#1}}}
\newcommand\MICHELE[1]{\textcolor{blue}{\nb{MICHELE}{#1}}}
\newcommand\ICE[1]{\textcolor{blue}{\nb{ICE}{#1}}}
\newcommand\BIN[1]{\textcolor{blue}{\nb{BIN}{#1}}}
\newcommand\CSABA[1]{\textcolor{blue}{\nb{CSABA}{#1}}}
\newcommand\HIDEAKI[1]{\textcolor{blue}{\nb{HIDEAKI}{#1}}}
\newcommand\KENICHI[1]{\textcolor{blue}{\nb{KENICHI}{#1}}}

\begin{abstract}

\edit{Modern} software is developed under considerable time pressure, which implies that developers more often than not have to resort to compromises when it comes to code that is well written and code that just does the job. This has led over the past decades to the concept of ``technical debt'', a short-term hack that \edit{potentially} generates long-term maintenance problems. Self-admitted technical debt (SATD) is a particular form of technical debt: developers consciously perform the hack but also document it in the code by adding comments as a reminder (or as an admission of guilt). We focus on a specific type of SATD, namely ``On-hold'' SATD, in which developers document in their comments the need to halt an implementation task due to conditions outside of their scope of work (\eg an open issue must be closed before a function can be implemented). 

We present an approach, based on regular expressions and machine learning, which is able to detect issues referenced in code comments, and to automatically classify the detected instances as either {\em ``On-hold''} (the issue is referenced to indicate the need to wait for its resolution before completing a task), or as \emph{``cross-reference''}, (the issue is referenced to document the code, for example to explain the rationale behind an implementation choice). Our approach also mines the issue tracker of the projects to check if the On-hold SATD instances are ``superfluous'' and can be removed (\ie the referenced issue has been closed, but the SATD is still in the code). Our evaluation confirms that our approach can indeed identify relevant instances of On-hold SATD. We illustrate its usefulness by identifying superfluous On-hold SATD instances in open source projects as confirmed by the original developers.

\end{abstract}

\begin{IEEEkeywords}
Self-admitted technical debt, empirical software engineering, \edit{issue}
\end{IEEEkeywords}

\section{Introduction} \label{sec:introduction}

Technical debt (TD) was first mentioned as a concept by Cunningham close to 30 years ago \cite{10.1145/157709.157715}, when he wrote the following lines: \emph{``Shipping first time code is like going into debt. A little debt speeds development so long as it is paid back promptly [...] The danger occurs when the debt is not repaid. Every minute spent on not-quite-right code counts as interest on that debt. Entire engineering organizations can be brought to a stand-still under the debt.''} 

In simple words, TD is a short-term ``hack'' (often induced by industrial reality, which dictates that either time and/or money are short) with long-lasting consequences if not properly handled. Since developers naturally keep working on new parts and do not revisit something unless it is strictly necessary, very often TD results, in the long run, in low maintainability and poor performance \cite{6280547}.

Potdar and Shihab extended the concept of TD to the notion of self-admitted technical debt (SATD) \cite{10.1109/ICSME.2014.31}, performed intentionally by developers, but mentioned/admitted as comments in the source code. They found that SATD is present, depending on the system, in 2.4\% to over 30\% of the files and that only 26\%-63\% gets removed, \ie a non-SATD often remains in the code. Zampetti \etal furthermore found that 20\% - 50\% of the removals were accidental and are even unintended \cite{8595236}.

Maldonado and Shihab categorized SATD into 5 types: design debt, defect debt, documentation debt, requirement debt, and test debt~\cite{7332619}, with design debt and requirement debt being the most common ones. Xavier \etal also found that SATD not only manifests itself as comments in the source code, but is also present in issue reports \cite{xavier2020beyond}.

We focus on a particular type of SATD, first introduced by Maipradit \etal \cite{10.1007/s10664-020-09854-3}: ``On-hold SATD'', defined as self-admitted technical debt due to a waiting condition for an external event to happen before the technical debt can be removed. 
\edit{In particular, this paper focuses on On-hold SATD with references to issues.}

\begin{figure}[ht]
    \vspace{-0.1cm}
    \centering
    \begin{subfigure}{\linewidth}
        \includegraphics[width=\textwidth]{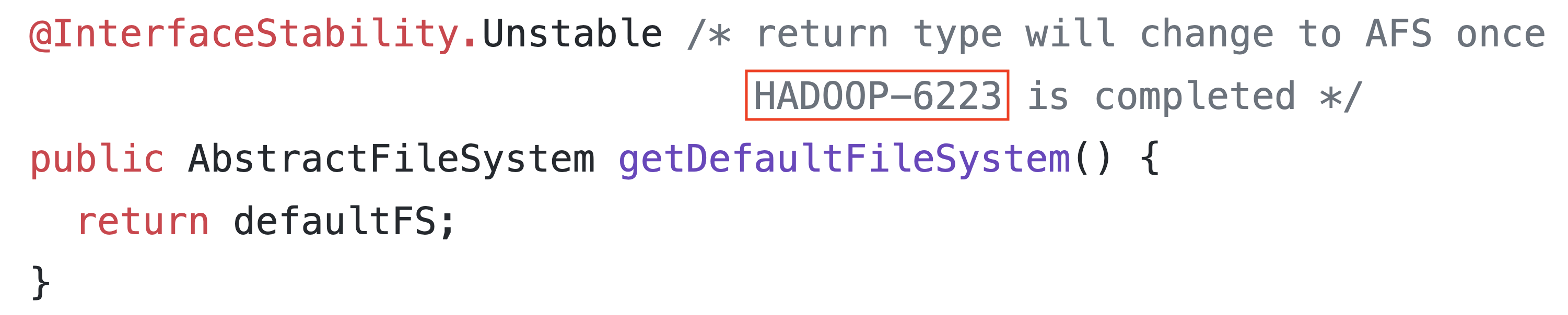}
        \caption{A SATD code comment referencing an issue}
    \end{subfigure}
    \vspace{0.3cm}
    
    \begin{subfigure}{\linewidth}
        \includegraphics[width=\textwidth]{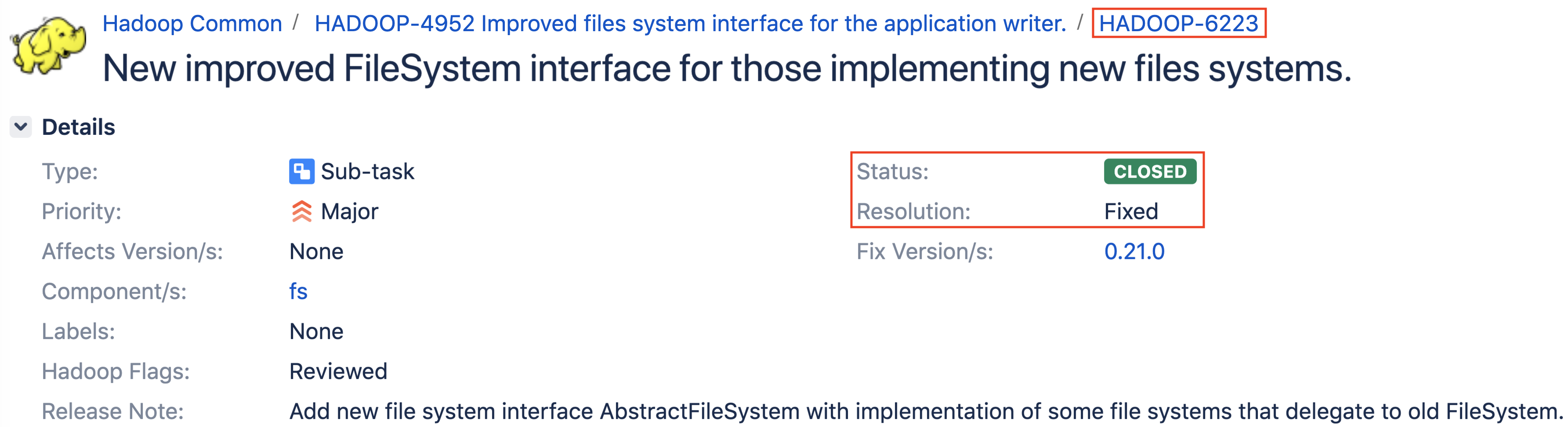}
        \caption{A referenced issue report (\url{https://tinyurl.com/ybunu2dj})}
    \end{subfigure}
    \caption{Motivating Example}
    \label{fig:motivating}
    \vspace{-0.1cm}
\end{figure}

{\bf Motivating Example.} \figref{fig:motivating}-(a) shows code from Apache Hadoop. The comment in the code indicates that an action will be taken once a condition is fulfilled, \ie the closing of issue 6223. As we see from \figref{fig:motivating}-(b), the issue has in fact already been closed, but the On-hold SATD was not removed, thus creating confusion to anyone inspecting the code. 

In essence, On-hold SATD are intentionally reminders left in the source code whose sole purpose is to be removed. 

We present a large-scale empirical study to ascertain whether (i) On-hold SATD can be automatically detected, and (ii) it is possible to identify cases in which the On-hold SATD should be removed, since the ``waiting condition'' has been fulfilled, thus making the SATD a form of ``wrong documentation'' in the code. Besides quantitatively evaluating the approaches we built to identify and remove On-hold SATD instances, we also show its usefulness in practice by collecting feedback from developers of open source projects.

\section{Related work} \label{sec:related_work}

\subsection{Empirical Studies on (self-admitted) Technical Debt}

Storey \etal \cite{Storey:icse2008} studied how annotations in code comments (\eg TODO, FIXME) are used by developers to keep track of tasks. Several types of activities are supported by these annotations, \eg the usage of TODOs to ask questions to other developers during code comprehension. These annotations are a subset of the ones used nowadays to detect SATD.

Guo \etal \cite{Guo:icsm2011} studied a specific technical debt instance to assessing its impact on the project costs. Their findings confirmed the harmfulness of technical debt, showing that the delayed task resulted in tripled implementation costs.

Klinger \etal \cite{Klinger:mtd2011} investigated how decisions to acquire technical debt are made within IBM by interviewing four technical architects. They found that technical debt is often due to imposed requirements to meet a specific deadline sacrificing quality. Also, the interviewed architects reported a lack of effective communication between technical and non-technical stakeholders involved in technical debt management.

Lim \etal \cite{Lim:ieee2012} interviewed practitioners (35 in this case) to investigate their perspective on TD. They found that most of the participants were familiar with the notion of TD and they do consider it as a poor programming practice, but more as an \emph{intentional decision to trade off competing concerns during development} \cite{Lim:ieee2012}. Practitioners also highlighted the difficulty in measuring the cost of TD. Similarly, Kruchten \etal \cite{Kruchten:2013} reported their understanding of the technical debt in industry as the result of a four-year interaction with practitioners.

Spinola \etal \cite{Spinola:2013} asked 37 practitioners to validate 14 statements about TD (\eg ``\emph{The root cause of most technical debt is pressure from the customer}'' \cite{Rubin:2012}). The statement achieving the highest agreement was ``\emph{If technical debt is not managed effectively, maintenance costs will increase at a rate that will eventually outrun the value it delivers to customers}''.

Kruchten \etal \cite{Kruchten:ieee2012} provided theoretical foundations to the concept of TD by presenting the ``technical debt landscape'', classifying TD as visible or invisible and highlighting the debt types causing evolvability and maintainability issues. Alves \etal \cite{Alves:MTD2014} proposed an ontology of terms on technical debt.

Potdar and Shihab \cite{10.1109/ICSME.2014.31} introduced the notion of SATD by mining five software systems to investigate (i) the amount of SATD they contain, (ii) the factors promoting the introduction of the SATD, and (iii) how likely is the SATD to be removed. Bavota and Russo \cite{Bavota:msr2016} performed a differentiated replication of that study involving a larger set of subject systems (159), confirming the findings of the original study.

Zazworka \etal \cite{Zazworka:ease2013} studied the overlap between the technical debt instances detected by automated tools and by manual inspection, finding very little overlap. 

Maldonado and Shihab \cite{Maldonado:MTD2015} used the TD classification by Alves \etal \cite{Alves:MTD2014} to investigate the types of SATD more diffused in open source projects. They identified 33k comments in five software systems reporting SATD. These comments have been manually read by one of the authors who found as the vast majority of them ($\sim$60\%) reported design debt. 

Wehaibi \etal~\cite{7476641} studied the relationship between SATD and software quality, finding that files with SATD do not have more defects compared to files without SATD, but that changes in the context of SATD are more complex. Sierra \etal~\cite{SIERRA201970} conducted a survey about SATD research, categorizing it into: detection, comprehension, and repayment. They found a lack of research related to repayment and management of SATD.

\subsection{Automatic detection/management of SATD}

In the study by Potdar and Shihab~\cite{10.1109/ICSME.2014.31}, the authors identified SATD using 62 textual patterns. The patterns can be matched in code comments of a previously unseen project to identify SATD. Farias \etal~\cite{7332621} built on top of these 62 patterns and developed a model called CVM-TD (Contextualized Vocabulary Model for identifying TD) that exploits combinations of the patterns to identify different types of technical debt. 

Maldonado \etal~\cite{7820211} presented an approach to automatically identify design and requirement SATD by applying Natural Language Processing (NLP) on code comments. A study performed on ten open source projects showed the superiority of their approach as compared to the state-of-the-art, represented at that time by the above-described pattern-based techniques. Wattanakriengkrai \etal~\cite{8661216} developed a classifier to identify design and requirements SATD using N-gram IDF and automated machine learning on Maldonado's dataset. Comparing the result with the previous study~\cite{7820211}, the classifier outperforms the NLP approach in both design and requirement. A similar idea has also been exploited by Huang \etal~\cite{Huang:2018:IST:3188697.3188709} that leveraged text-mining for SATD identification. Also in this case, the approach performed better than the pattern-based approach by Potdar and Shihab~\cite{10.1109/ICSME.2014.31}. This approach is also available as an Eclipse plug-in \cite{8449432}.

Ren \etal~\cite{10.1145/3324916} proposed an approach based on Convolution Neural Networks to classify code comments into SATD or non-SATD. An experiment performed on ten projects and 63k comments showed that their approach outperforms text mining techniques both for within-project and cross-project prediction. 

Zampetti \etal~\cite{8094423} presented TEDIOUS (TEchnical Debt IdentificatiOn System), an approach to train a recommender to suggest developers writing new code when to self-admit design TD, or improve the code being written. TEDIOUS achieves an average precision of $\sim$50\%. Yan \etal~\cite{8352718} proposed a model to determine whether a change introduces SATD. They manually labeled changes that introduced SATD in the past and built a model exploiting 25 features to characterize SATD-introducing changes. An empirical study across $\sim$100k changes reported an AUC for the model of 0.82.

\begin{table*}[ht]
    \caption{Details of the projects in our dataset. SLOC is calculated on Java files using SLOCCounts~\cite{wheeler2004sloc}.}
    \centering
    \label{tab:project_detail}
    \begin{tabular}{lllrrrr}
        \toprule
        & & & & & \textbf{\# \edit{Remaining} comments} & \textbf{\# Removed comments} \\
        \textbf{Project} & \textbf{Version} & \textbf{ITS} & \textbf{SLOC} & \textbf{\# Contributors} &  \textbf{that refer to issues} & \textbf{that refer to issues} \\
        \midrule
        Apache Ant & 1.10.7 & Bugzilla & 144,966 & 47 & 27 & 22 \\
        Apache Camel & 3.0.0 & Jira & 1,267,905 & 544 & 42 & 62 \\
        Apache Dubbo & 2.7.4 & Github & 148,377 & 268 & 8 & 4 \\
        Apache Hadoop & 2.10.0 & Jira & 1,885,604 & 239 & 272 & 269 \\
        Apache Jmeter & 5.2.1 & Bugzilla & 142,030 & 19 & 116 & 136 \\
        Apacha Kafka & 2.4.0 & Jira & 319,990 & 606 & 24 & 21 \\
        Apache Log4j & 1.2.17 & Bugzilla & 30,608 & 7 & 6 & 3 \\
        Apache Logging-log4j2 & 2.13.0 & Jira & 159,353 & 76 & 179 & 153 \\
        Apache Tomcat & 10.0.0 & Bugzilla & 341,192 & 31 & 82 & 73 \\
        Mockito & 3.3.10 & Github & 48,292 & 173 & 15 & 16 \\
        \midrule
        Total & - & - & 4,488,317 & 2,010 & 771 & 759 \\
        \bottomrule
    \end{tabular}
\end{table*}

\begin{table*}[ht]
    \centering
    \caption{Regular expressions to identify issue in comments}
    \label{regex}
    \begin{tabular}{llll}
    \toprule
    \textbf{ITS} & \textbf{Regular expression} \\
    \midrule
    Bugzilla & \texttt{(?<![A-Za-z])(?:bug|projectname|bugzilla|bz)[ -](?:\#)?\textbackslash d+(?:\textbackslash .[0-9xX*]+)*} \\
     & \# issue IDs, e.g., Bug 34383 \\
     & \texttt{https?:\textbackslash/\textbackslash/[\textbackslash w.\_/]*show\_bug.cgi\textbackslash ?id=\textbackslash d+} \\
     & \# URLs, e.g., \texttt{https://bz.apache.org/bugzilla/show\_bug.cgi?id=51687} \\
    \midrule
    Github & \texttt{(?<![A-Za-z])(?:bug|issues?)[ -](?:\#)?\textbackslash d+(?:\textbackslash .[0-9xX*]+)*} \\
    & \# issue IDs, e.g., issue 55 \\
    & \texttt{https?:\textbackslash/\textbackslash/github.com/[\textbackslash w.\_/]*\textbackslash/issues\textbackslash/\textbackslash d+} \\
    & \# URLs, e.g., \texttt{https://github.com/apache/dubbo/issues/3251} \\
    \midrule
    Jira & \texttt{(?<![A-Za-z])(?:bug|projectname)[ -](?:\#)?\textbackslash d+(?:\textbackslash .[0-9xX*]+)*} \\
    & \# issue IDs, e.g., HADOOP-7234 \\
    \bottomrule
    \end{tabular}
\end{table*}

\section{Approach} \label{sec:approach}

\begin{figure}[ht]
    \centering
    \includegraphics[width=0.8\linewidth]{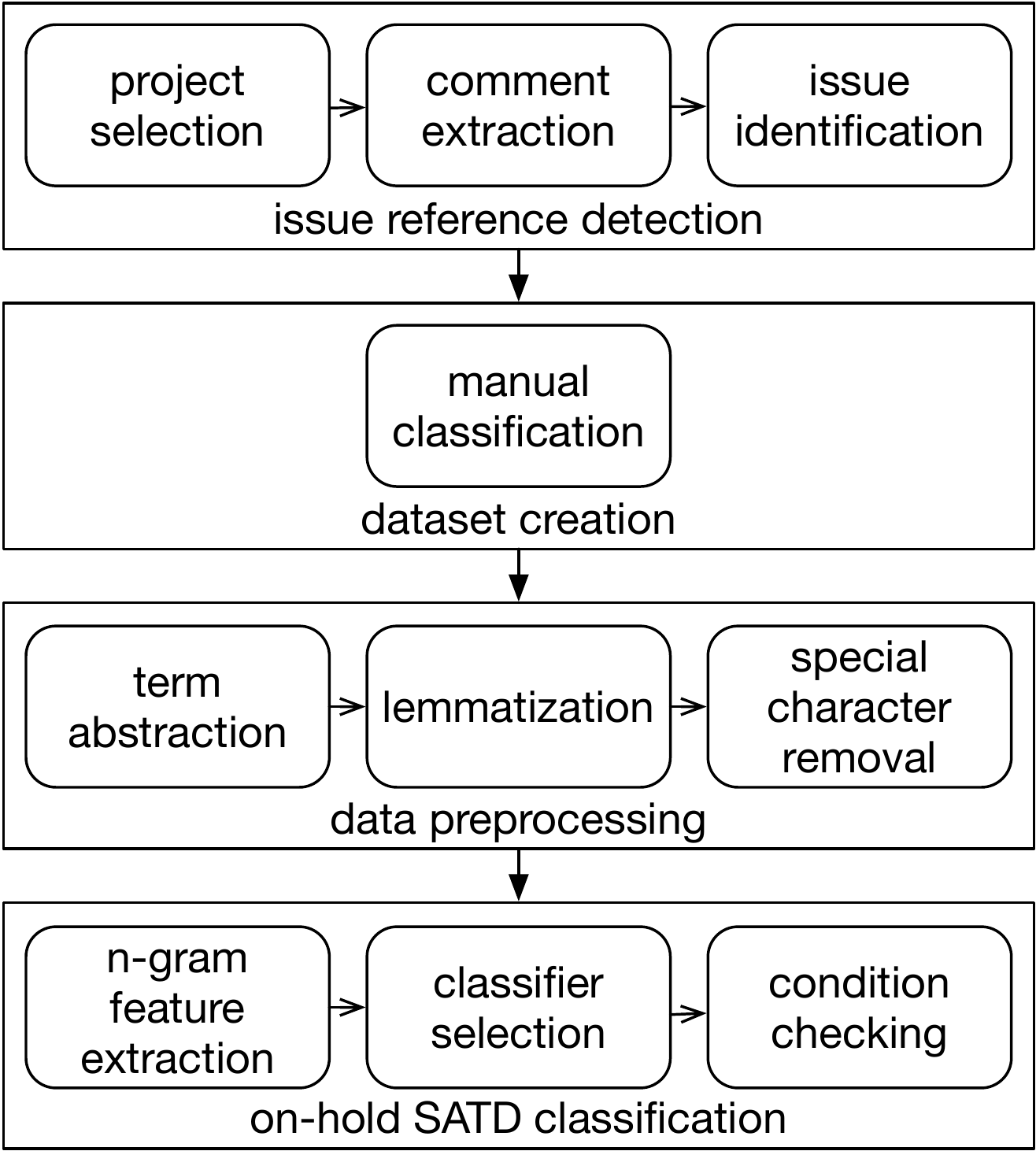}
    \caption{Approach for On-hold SATD detection and removal}
    \label{fig:approach}
\end{figure}

We aim to build a classifier which automatically detects On-hold SATD and indicates whether it is ready to be removed. To achieve this goal, we took the following four steps (\figref{fig:approach}): 1) issue reference detection, 2) dataset creation, 3) data preprocessing, and 4) On-hold SATD classification.

\subsection{Issue Reference Detection}

To detect On-hold SATD, our first step is to locate the code comments referring to issues.

\subsubsection{Project Selection}

We selected ten open source projects that consistently used for their entire change history a specific issue tracking system (ITS). This allowed us to run our study without the risk of missing important information due to the migration between different issue tracker systems (\eg starting on JIRA and then moving to the GitHub issue tracker). \tabref{tab:project_detail} lists the projects used in this study.

\subsubsection{Comment Extraction}

We iterated over the commits in the repositories of the selected projects, and extracted all single line comments (\eg~// ...) and multi-line comments (\eg~/* ... */) from Java files. If multiple comments are next to each other (\eg~/* ... */ // ...) they are considered as a single block of comments. Comments from test files are ignored, as issue references there are most likely to serve as explanations of what developers are testing, instead of SATD.

\subsubsection{Issue Identification}

Issue references are identified using regular expressions to match issue IDs and issue URLs. \tabref{regex} shows the regular expressions used for each issue tracking system. For each identified issue reference, we also recorded its life cycle: We iterated over the commit history and extracted the date when the issue reference was first introduced in the comment, and in case, when it was removed.

From the 10 selected projects, we identified 1,530 comments containing issue references, among which 759 had already been removed, while the remaining 771 still \edit{remain} in the latest commit by the date of data collection.

\begin{table*}[ht]
    \centering
    \caption{Regular expressions for term abstraction}
    \label{tab:term_abstraction}
    \begin{tabular}{lll}
    \toprule
    \textbf{String} & \textbf{ITS} & \textbf{Regular expression} \\
    \midrule
    abstractissueid & Bugzilla & \texttt{(?<![A-Za-z])(?:bug|projectname|bugzilla|bz)[ -](?:\#)?\textbackslash d+(?:\textbackslash .[0-9xX*]+)*} \\
    & & \# issue IDs \\
    & & \texttt{https?:\textbackslash/\textbackslash/[\textbackslash w.\_/]*show\_bug.cgi\textbackslash ?id=\textbackslash d+} \\
    & & \# URLs \\
    \cline{2-3}
    & Github & \texttt{(?<![A-Za-z])(?:bug|issues?)[ -](?:\#)?\textbackslash d+(?:\textbackslash .[0-9xX*]+)*} \\
    & & \# issue IDs \\
    & & \texttt{https?:\textbackslash/\textbackslash/github.com/[\textbackslash w.\_/]*\textbackslash/issues\textbackslash/\textbackslash d+} \\
    & & \# URLs \\
    \cline{2-3}
    & Jira & \texttt{(?<![A-Za-z])(?:bug|projectname)[ -](?:\#)?\textbackslash d+(?:\textbackslash .[0-9xX*]+)*} \\
    & & \# issue IDs \\
    & & \texttt{https?:\textbackslash/\textbackslash/issues.apache.org/\textbackslash/jira\textbackslash/browse\textbackslash/(?:projectname)-\textbackslash d+} \\
    & & \# URLs \\
    \midrule
    abstracturl & --- & \texttt{https?:\textbackslash/\textbackslash/(www\textbackslash.)?[-a-zA-Z0-9@:\%.\_\textbackslash+\textasciitilde\#=]\{2,256\}\textbackslash.[a-z]\{2,6\}\textbackslash b} \\
    & & \texttt{([-a-zA-Z0-9@:\%\_\textbackslash+.\textasciitilde\#?\&//=]*)}\\
    \bottomrule
    \end{tabular}
\end{table*}

\subsection{Dataset Creation} \label{sec:data_creation}

To build the On-hold SATD classifier, we created a dataset for training and testing, based on the issue-referring comments collected in our previous step. For each of the 1,530 comments, the first and the second author independently labeled whether it is an actual instance of On-hold SATD or, instead, it is used as cross-reference. We evaluated the inter-rater reliability with the Cohen's kappa coefficient, and the score of 0.748 demonstrates a substantial agreement between the two labelers. The third author resolved labeling conflicts. As a result, we got 133 On-hold SATD and 1,397 cross-reference comments.

\tabref{tab:dataset} summarizes the annotation results. 133 (8.7\%) of the issue-referring comments are instances of On-hold SATD. 

\begin{table}[h]
    \centering
    \caption{Statistics of annotated comments containing issue references}
    \label{tab:dataset}
    \begin{tabular}{lrrr}
        \toprule
        & \textbf{On-hold SATD} & \textbf{Cross-reference}  & \textbf{Total} \\
        \midrule
       \edit{Remaining} comments & 40 & 731 & 771 \\
       Removed comments & 93 & 666 & 759 \\
       \midrule
       \edit{Total} & 133 & 1,397  & 1,530  \\
       \bottomrule
       \end{tabular}       
\end{table}

\subsection{Data Preprocessing}

Before extracting features from the comments and feeding them into the classifier, we performed three preprocessing steps: 1) term abstraction, 2) lemmatization, and 3) special character removal.

\subsubsection{Term Abstraction}

For all the comments, we abstracted issue IDs and hyperlinks referring to issues to the string ``\texttt{abstractissueid}'', while the hyperlinks unrelated to issues were abstracted to ``\texttt{abstracturl}''. This is done to eliminate the impact of issue IDs and hyperlinks during classification, as we are not interested in their real content. \tabref{tab:term_abstraction} summarizes the regular expressions we used to extract relevant issue IDs and hyperlinks for different issue tracking systems.

\subsubsection{Lemmatization}

We applied lemmatization with the Spacy natural language processing tool \cite{spacy2}, which normalizes words with the same root but different surfaces into the same format \cite{keselj2009speech}. For example, the words ``sang'', ``singing'', and ``sings'' will be converted into ``sing''.

\subsubsection{Special character removal}

We removed all non-English and non-numeric characters using the regular expression \texttt{[\textasciicircum A-Za-z0-9]+}.

For our study we did not apply stop word removal, a commonly used text preprocessing step, as it might remove some keywords important for identifying On-hold SATD, such as ``when'' and ``until''.

\subsection{On-hold SATD Classification}

After preprocessing, we extracted n-gram features from the comments and used them to train a classifier to identify On-hold SATD. We also checked issue status and issue resolution to determine whether an On-hold SATD comment is ready to be removed.

\subsubsection{N-gram Feature Extraction}

Similar to another SATD classification approach by Wattanakriengkrai \etal \cite{8661216}, we extracted n-gram features by applying n-gram IDF~\cite{10.1145/3052775,10.1145/2736277.2741628}. N-gram IDF is a theoretical extension of IDF (Inverse Document Frequency). The traditional IDF approach assigns more weight to terms occurring in fewer documents, which does not work well for n-grams. For example, ``Leonardo da is'' might have higher weight than ``Leonardo da Vinci''. N-gram IDF is designed to address this issue and can determine the dominant n-grams and extract key terms of any length~\cite{10.1145/3052775,10.1145/2736277.2741628}.
In this study, we extracted n-grams from SATD comments using the library n-gram weighting scheme~\cite{iwnsew_2017} with default settings. We obtained the list of all valid n-gram terms containing up to 10-gram terms. \edit{In total, we receive 644 terms of n-grams.}

\subsubsection{Classifier Selection}

After extracting the n-gram terms, we build a classifier to identify bug referencing comments into On-hold SATD or not. While there many different algorithms available for supervised classification, it is hard to decide which one to pick, as different datasets and hyper-parameter settings might both impact the performance of these algorithms. Automated machine learning addresses this problem by running multiple classifiers with different parameters to optimize performance. In this study, we used auto-sklearn~ \cite{NIPS2015_5872}, which includes 15 classification algorithms, 14 feature preprocessing and 4 data preprocessing techniques~\cite{NIPS2015_5872}.

\subsubsection{Condition Checking} \label{sec:condition_check}

After identifying the On-hold SATD using our classifier, \edit{our program automatically checks} the referred issue status and resolution to decide whether the SATD is ready to be removed. In the issue tracking system, if the status of the referred issue is set to ``resolved'', ``closed'', or ``verified'', and the field of resolution (if applicable) is set to ``fixed'', we consider it ready for removal.

\section{Study Design}

The \emph{goal} of this study is to evaluate the accuracy of our approach for On-hold SATD identification and removal. Moreover, we are interested in the evolution of On-hold SATD in open source projects. The \emph{context} of the study consists of 1,530 code comments containing issue references, extracted from the previously presented 10 open source projects. 

\subsection{Research Questions}

In this study, we answer the following three research questions (RQs):

\begin{itemize}

    \item \textbf{RQ$_1$:} \emph{What is the accuracy of our approach in identifying On-hold SATD?} This RQ investigates the performance of our classifier in identifying On-hold SATD. We also examined the impact of oversampling, different features and machine learning algorithms on the performance of our classifier:   
    
    \begin{itemize}
        \item \textbf{RQ$_{1.1}$:} \emph{How do n-grams impact the performance of our classifier as compared to Bag-Of-Words features?}
        \item \textbf{RQ$_{1.2}$:} \emph{How does oversampling impact the performance of the classifier?}
        \item \textbf{RQ$_{1.3}$:} \emph{How do different machine learning algorithms impact the performance of the classifier?}
    \end{itemize}

    \item \textbf{RQ$_2$:} \emph{How does On-hold SATD evolve in open source projects?} To gain deeper insights on how On-hold SATD evolves in the projects, with this RQ we inspect the duration of existence of On-hold SATD in software projects, and the time it takes to address SATD after the relevant issue is resolved.

    \item \textbf{RQ$_3$:} \emph{To what extent can our approach identify ``ready-to-be-removed'' On-hold SATD?} This RQ empirically evaluates the reliability of our approach in identifying On-hold SATD which should be removed, since it was already ``paid back''.
    
\end{itemize}
    
\subsection{Context Selection \& Data Collection}

In this study, we used the dataset presented in \secref{sec:data_creation}, which contains 1,530 annotated comments containing issue references. 

To answer RQ$_1$, we built a classifier using auto-sklearn with n-grams extracted by n-gram IDF \cite{8661216} as features. N-grams were extracted from On-hold SATD comments only. N-grams from Cross-reference comments are not included because we want to extract important patterns to detect on-hold SATD, and we use these patterns to discriminate between On-hold SATD and Cross-reference. We performed a ten-fold cross validation: We divided the 1,530 issue-referring comments into ten different sets, each one composed of 153 comments. Then, we iteratively used one set as the \emph{test set}, while the remaining 1,377 comments were used for \emph{training}.
 
To answer RQ$_{1.1}$, we ran a different classifier implementation on the dataset, using Bag-Of-Words (BOW) as features.

To answer RQ$_{1.2}$, we applied an oversampling technique (\ie SMOTE) to our training set, and then compared the results achieved by our classifier with/without oversampling.

To answer RQ$_{1.3}$, we built three variants of the classifier with different machine learning algorithms: Naive Bayes, Support Vector Machine (SVM), and K-Nearest Neighbors (KNN). 

To answer RQ$_2$, we inspected the removed issue-referring comments for both On-hold and cross-reference comments. We first checked the time interval between the introduction and the removal of these comments. Then, for the instances referring issues that have been solved, we compute the difference between the issue resolution time and the corresponding On-hold SATD removal event.

\begin{figure}[ht]
    \centering
    \includegraphics[width=\linewidth]{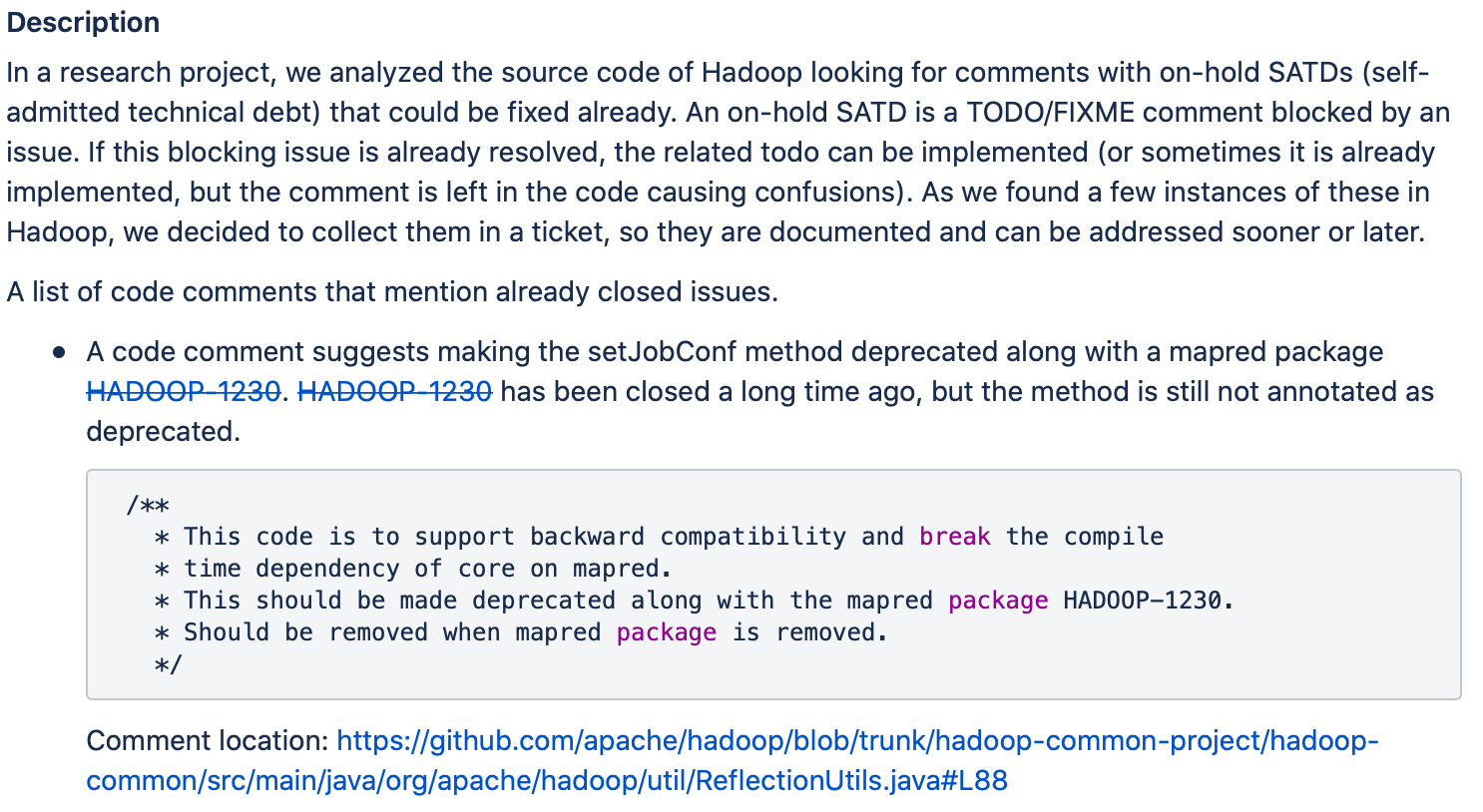}
    \caption{An example issue report.}
    \label{fig:issue_report}
\end{figure}

To answer RQ$_3$, we identified the On-hold SATD comments which are ready to be removed from the \edit{40 still remaining On-hold} issue-referring comments, based on the corresponding issue status and resolution, as described in \secref{sec:condition_check}. In total, we identified 10 ``ready-to-be-removed'' On-hold SATD comments. By the time we started working on RQ$_3$, 4 of 10 comments had already been removed by developers (three were removed thanks to code changes addressing the On-hold SATD, while one was removed due to the deletion of the file containing it). We reported the remaining six ``ready-to-be-removed'' On-hold SATD comments to the developers by creating issue reports in the respective issue tracker. In the issue report, we inform developers why the On-hold SATD comments should be removed and where they are located. An example of the issue reports can be seen in \figref{fig:issue_report}.

\begin{table*}[ht]
    \centering
    \caption{Performance of classifiers in identifying On-hold SATD}
    \label{tab:performance}
    \begin{tabular}{@{}l|r|r|r|rrr@{}}
        \toprule
        & \multicolumn{1}{c|}{\textbf{Original approach}}      & \multicolumn{1}{c|}{\textbf{BOW as feature}}   & \multicolumn{1}{c|}{\textbf{With Oversampling}}   & \multicolumn{3}{c}{\textbf{Different ML algorithms}}  \\ \midrule
    \multicolumn{1}{l|}{} & \multicolumn{1}{c|}{\begin{tabular}[c]{@{}c@{}}n-gram + \\ auto-sklearn\end{tabular}} & \multicolumn{1}{c|}{\begin{tabular}[c]{@{}c@{}}BOW + \\ auto-sklearn\end{tabular}} & \multicolumn{1}{c|}{\begin{tabular}[c]{@{}c@{}}n-gram + oversampling +\\ auto-sklearn\end{tabular}} & \multicolumn{1}{c}{\begin{tabular}[c]{@{}c@{}}n-gram + \\ Naive Bayes\end{tabular}} & \multicolumn{1}{c}{\begin{tabular}[c]{@{}c@{}}n-gram + \\ SVM\end{tabular}} & \multicolumn{1}{c}{\begin{tabular}[c]{@{}c@{}}n-gram + \\ KNN\end{tabular}} \\ 
        \midrule
        Precision & 0.79 & 0.69 & 0.38 & 0.64 & 0.87 & 0.88 \\
        Recall & 0.70 & 0.68 & 0.48 & 0.56 & 0.38 & 0.15 \\
        F1-score & 0.73 & 0.67 & 0.41 & 0.59 & 0.51 & 0.25 \\
        AUC & 0.97 & 0.94 & 0.87 & 0.81 & 0.95 & 0.76 \\ \bottomrule
    \end{tabular}
\end{table*}

\begin{table*}[ht]
    \centering
    \caption{Statistical results of performance comparisons of classifiers}
    \label{tab:stat_performance}
    \begin{tabular}{@{}lrrrr@{}} \toprule
      & \textbf{P-value (Precision)} & \textbf{Effect size (Precision)}  & \textbf{P-value (Recall)} & \textbf{Effect size (Recall)}  \\ \midrule
    \emph{n-gram+auto-sklearn} vs \emph{BOW+auto-sklearn}                   &   $<$ 0.01      &   0.48 (large)  &   0.32       &  -     \\ \midrule
    \emph{n-gram+auto-sklearn} vs \emph{n-gram+oversampling+auto-sklearn} &    $<$ 0.01       &  0.92 (large)  &    0.01   &   0.67 (large) \\ \midrule 
    
    \emph{n-gram+auto-sklearn} vs \emph{n-gram+Naive Bayes}                   &   0.06      & - &   0.03    &    0.58(large)    \\ 
    \emph{n-gram+auto-sklearn} vs \emph{n-gram+SVM} &  0.30   &  -   &     0.03    &    0.8 (large)    \\
    \emph{n-gram+auto-sklearn} vs \emph{n-gram+KNN}                   &  0.30    &   -  &     0.03    &  1.0 (large)    \\ 
    \emph{n-gram+Naive Bayes} vs \emph{n-gram+SVM} &   0.06      &  -   &  0.03   &     0.58 (large)   \\
    \emph{n-gram+Naive Bayes} vs \emph{n-gram+KNN}                   &   0.30      &  -   &   0.03      &   1.0 (large)    \\ 
    \emph{n-gram+SVM} vs \emph{n-gram+KNN} &    0.31     &   -  &   0.03  &   0.74 (large)   \\
    \bottomrule

    \end{tabular}
\end{table*}

\subsection{Data Analysis}

To answer \textbf{RQ$_{1}$} we compare the precision, recall, F1-score, and area under the ROC curve (AUC) of each experimented approach in classifying issue-referring comments (as belonging or not to On-hold SATD) for the dataset of 1,530 comments.

The comparisons are also performed via the Mann-Whitney test \cite{Cono1999a}, with results intended as statistically significant at $\alpha = 0.05$. For RQ$_{1.3}$, to control the impact of multiple pairwise comparisons (\eg the precision of auto-sklearn is compared with Naive Bayes, SVM, and KNN), we adjust $p$-values with Holm's correction \cite{Holm1979a}. We estimate the magnitude of the differences by using the Cliff's Delta ($d$), a non-parametric effect size measure \cite{Gris2005a}. We follow well-established guidelines to interpret the effect size: negligible for $|d| < 0.10$, small for $0.10 \le |d| < 0.33$, medium for $0.33 \le |d| < 0.474$, and large for $|d| \ge 0.474$ \cite{Gris2005a}.

To answer \textbf{RQ$_{2}$}, we present via violin plots the life spans of both On-hold SATD and cross-reference comments, as well as the duration between the resolution of issues and the removal of corresponding SATD comments. 

To answer \textbf{RQ$_{3}$}, we qualitatively analyze the developers' feedback.

\section{Results}

\subsection{RQ$_1$: What is the accuracy of our approach in identifying On-hold SATD?}

\begin{figure*}[ht]
    \centering
    \includegraphics[width=0.825\linewidth]{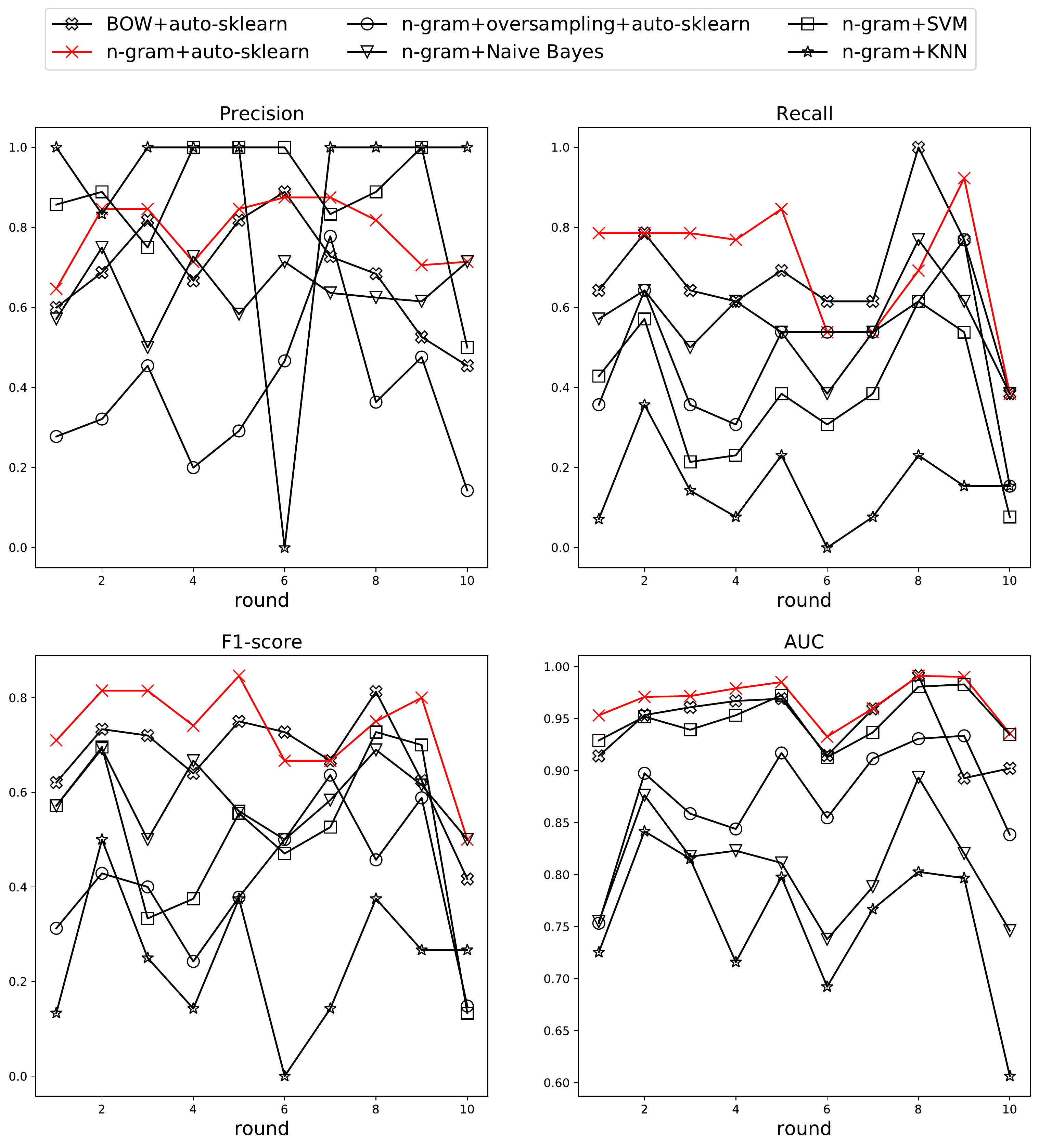}
    \caption{Results of each round in 10-fold evaluation for different classifier implementations}
    \label{fig:round}
\end{figure*}

\tabref{tab:performance} reports the average precision, recall, F1-score, and AUC of each experimented classifier implementations during 10-fold evaluation.

\tabref{tab:stat_performance} reports the statistical results of comparisons between different classifier implementations.

\figref{fig:round} also shows the results of the 10-fold evaluation for each experimented classifier in terms of precision, recall, F1-Score, and AUC.

As can be seen from \tabref{tab:performance}, the precision, recall, and F1-score achieved by our approach (``n-gram + auto-sklearn'') are all between 0.7 to 0.8, while AUC is as high as 0.97. This result demonstrates the reliability of our approach in On-hold SATD detection. 

To gain a better understanding of how our classifier works, we list the important n-gram features which frequently appear in On-hold SATD comments %
 in \tabref{tab:ngram_feature}. 

\begin{table}[ht]
    \centering
    \caption{N-gram features which frequently appear in On-hold SATD comments}
    \begin{tabular}{lr}
        \toprule
        \textbf{N-gram features} & \textbf{Frequency}\\
        \midrule
        `after', `abstractissueid' & 20 \\
        `once', `abstractissueid' & 18 \\
        `for', `now' & 12 \\
        `temporary', `fix' & 10 \\
        `workaround' & 10 \\
        `this', `be', `a', `temporary' & 8 \\
        `via', `abstractissueid' & 7 \\
        `be', `commit' & 7 \\
        `can', `be', `remove' & 7 \\
        `remove', `after', `abstractissueid' & 5 \\
        \bottomrule
    \end{tabular}
    \label{tab:ngram_feature}
\end{table}

These features help discriminate On-hold SATD from cross-reference. We can see that n-grams such as ``\emph{once abstractissueid}'', ``\emph{this be a temporary}'', and ``\emph{remove after abstractissueid}'' are especially important for identifying On-hold SATD. 

\begin{table*}[ht]
    \centering
    \caption{Example of classification results of our approach}
    \label{tab:example}
    \begin{tabular}{@{}cl@{}}
        \toprule
    \multicolumn{1}{l}{\textbf{Type}} & \textbf{Comment}         \\ \midrule
    True                                               & TODO: \textbf{workaround} (filling fixed bytes), to \textbf{remove after HADOOP-11938}       \\ \cmidrule{2-2} 
    Positive                                           &  ... This is a \textbf{temporary fix} ... See the discussion on HDFS-1965.   \\ \midrule
    False                                              & TODO: Temporarily keeping ... This has \textit{to be revisited} as part of HDFS-11029.         \\ \cmidrule{2-2} 
    Negative                                           &\textit{placeholder for} javadoc to prevent broken links, \textit{until HADOOP-6920}       \\ \midrule
    False                                           & \textbf{TODO: after MAPREDUCE-2793} YarnException is probably not expected here anymore but keeping it for now ... \\ \cmidrule{2-2} 
    Positive                                           & ... (CAMEL-9657) \textbf{[TODO] Remove} in 3.                                    \\ \bottomrule
    \end{tabular}
    \end{table*}

Additionally, we also illustrate some classification results in \tabref{tab:example}. From the two true positive examples (correctly identified as On-hold SATD by our approach), we can clearly see the patterns including ``\emph{workaround}'', ``\emph{temporary fix}'' and ``\emph{remove after abstractissueid}'', which can be related to  \tabref{tab:ngram_feature}. Therefore, it is not surprising that our classifier can correctly identify these On-hold SATD comments.

If we take a look at the two false negative examples (On-hold SATD classified as cross-reference), we find that phrases like ``\emph{to be revisit}'' and ``\emph{until abstractissueid}'' are probably useful n-grams for identifying On-hold SATD. Due to absence or infrequent occurrence, these n-grams are not used as features for the classifier. Expanding the training set can be a potential way for addressing the n-gram feature limitations.

In the two false positive examples (cross-reference classified as On-hold SATD), we can see that the n-gram terms like ``\emph{todo after abstractissueid}'' and ``\emph{todo remove}'' can be actually matched, and our classifier misclassified them into On-hold SATD. However, if we check the comments carefully, we can find that in the first sentence, it is clear that the issue has already been resolved, however, for some reason the developers decided to say ``\emph{keeping it for now}'', where ``\emph{it}'' refers to \texttt{YarnException}. In the second sentence, what follows ``\emph{todo remove}'' is actually not a reference to an issue, but a reference to a version. Some heuristic rules might help our classifier to better deal with these cases.

To understand how n-grams impact the performance of our classifier as compared to Bag-Of-Words (BOW) features, we inspect the first two columns of \tabref{tab:performance}, and the first row of \tabref{tab:stat_performance}. Using n-grams as features leads to a higher precision with a statistically significant difference and a large effect size. As for the recall, while the average value is higher when using n-grams, the performed analysis does not indicate a statistically significant difference. We conclude that compared to BOW features, n-grams lead to a significantly higher precision.

To understand how oversampling impacts the performance of the classifier, we inspect the first and the third column of \tabref{tab:performance}, as well as the second row of \tabref{tab:stat_performance}. \newpage

From the tables we can observe that the classifier obtains a statistically significant higher precision and recall when oversampling is not applied. Meanwhile, the effect sizes for both precision and recall comparisons are large. Indeed, after applying oversampling, the average precision, recall, F1-score, and AUC drop by around 40\%, 20\%, 30\%, and 10\%, respectively. We conclude that oversampling reduces the performance of our classifier in identifying On-hold SATD. 

To understand how different machine learning algorithms impact the performance of the classifier, we inspect the first and the last three columns of \tabref{tab:performance}, as well as the last six rows of \tabref{tab:stat_performance}. From the tables we can see that all the implementations achieved comparable precisions (from 0.64 to 0.88). Indeed, there is no statistically significant difference in terms of precision among these implementations. However, the differences emerge when comparing recall. Using auto-sklearn achieves a significantly higher recall than classifiers using other machine learning algorithms (\ie Naive Bayes, SVM and KNN). 

We also inspected which machine learning algorithm was adopted by auto-sklearn after automatic classifier selection. The records show that in 9 of the ten rounds of 10-fold evaluation Extra Trees was adopted, while the remaining one adopted Random Forest. That is, these two machine learning algorithms would potentially be a good choice for identifying On-hold SATD when automatic selection of the classifier is not possible. 

\subsection{RQ$_2$: How does On-hold SATD evolve in open source projects?}

\begin{figure}[ht]
    \centering
    \includegraphics[width=\linewidth]{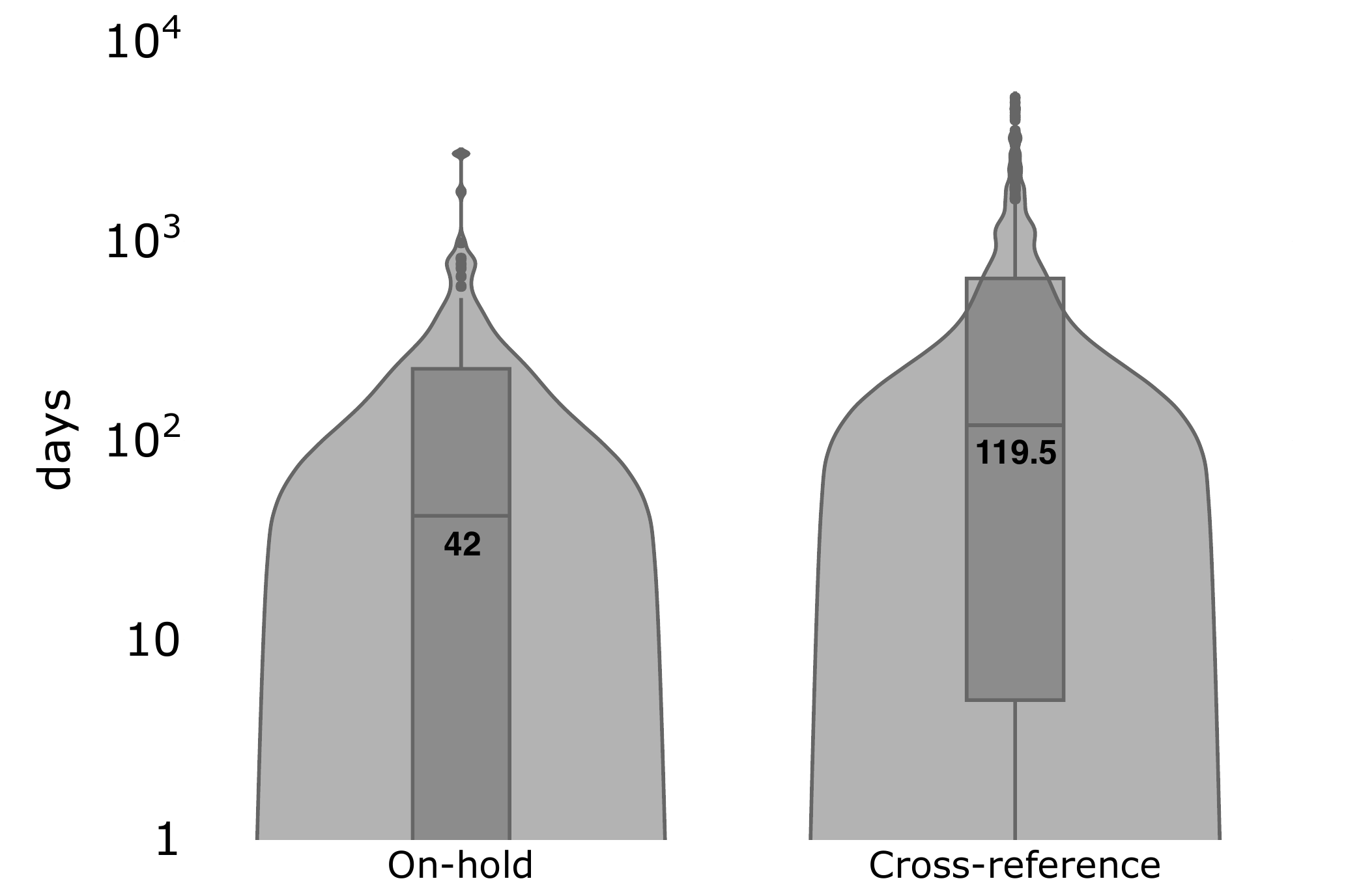}
    \caption{Distribution of life spans of removed issue-referring comments}
    \label{fig:violin}
\end{figure}

To answer RQ$_2$, we first looked into the life span of removed issue-referring comments for On-hold SATD and cross-reference comments separately. The life span distributions can be found in \figref{fig:violin}.

The median life span of On-hold SATD comments is 42 days, while it is 119.5 days for cross-reference comments. That is, overall, the median life span of cross-reference comments is almost three times of that of On-hold SATD.

Indeed, while both types of comments contain issue references, only On-hold SATD requires maintenance actions from developers. Cross-reference comments stay much longer as they are usually used for documentation purposes. 

\begin{figure}[ht]
    \centering
    \includegraphics[width=0.7\linewidth]{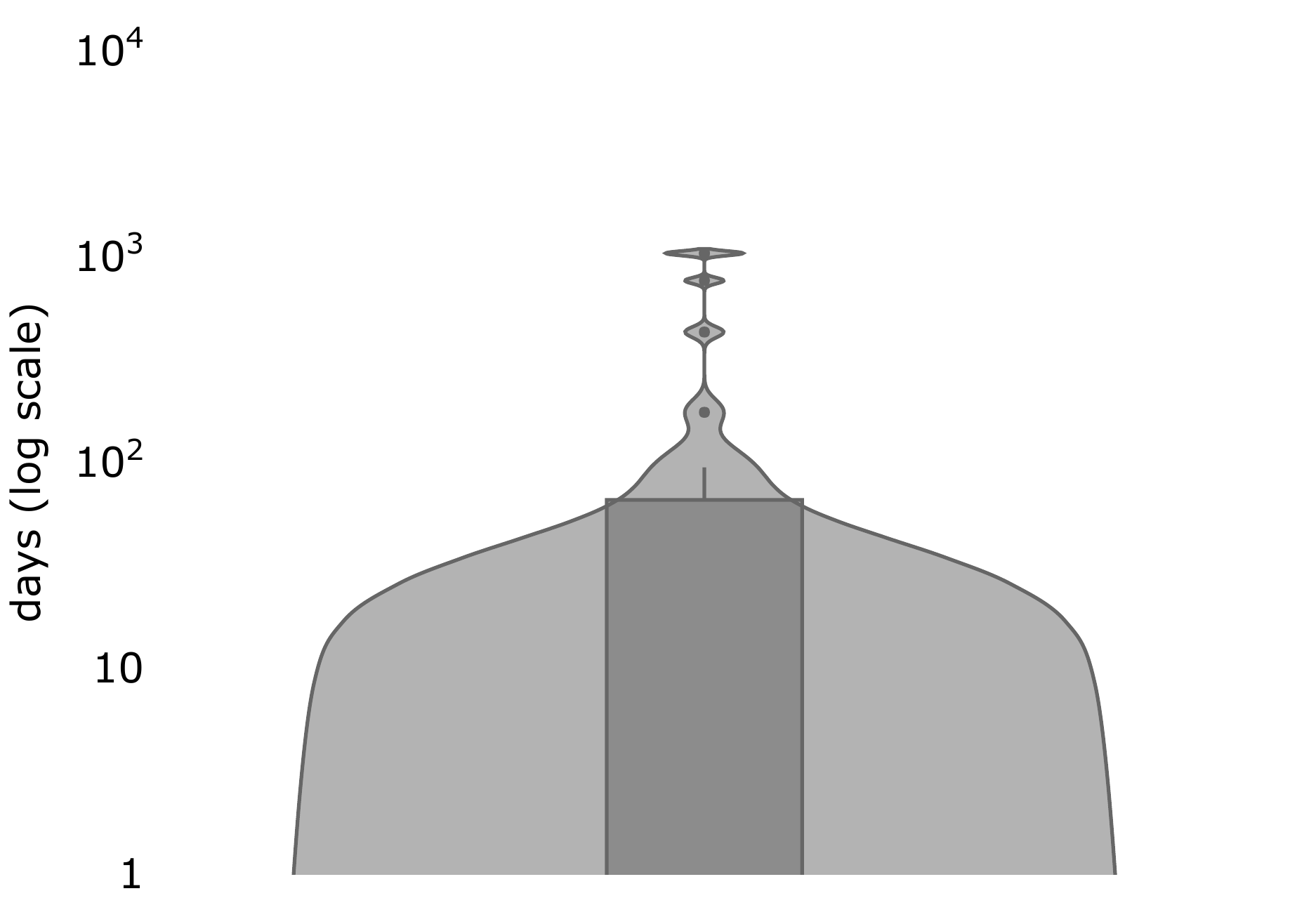}
    \caption{Distribution of days needed to address SATD comments after issues were resolved}
    \label{fig:distribution}
\end{figure}

We then investigated how long it takes to address On-hold SATD comments after the corresponding issues are resolved, and plotted the duration distribution in \figref{fig:distribution}. 

Around 53\% of On-hold SATD were removed within the same day when the issue was resolved. However, it takes longer than one year to remove 13\% of On-hold SATD. 

Additionally, we observed that some developers did not wait until the issue was resolved to address On-hold SATD comments. In fact, from a total of 93 removed On-hold SATD comments, we found that only 30 of them were removed after the issues were resolved. The corresponding issues of 9 On-hold SATD comments are still open or have the resolution set to ``wontfix''. 54 On-hold SATD comments were removed before the issues were resolved, although these issues have been resolved in the meantime.

\begin{figure*}[ht]
    \centering
    \includegraphics[width=\linewidth]{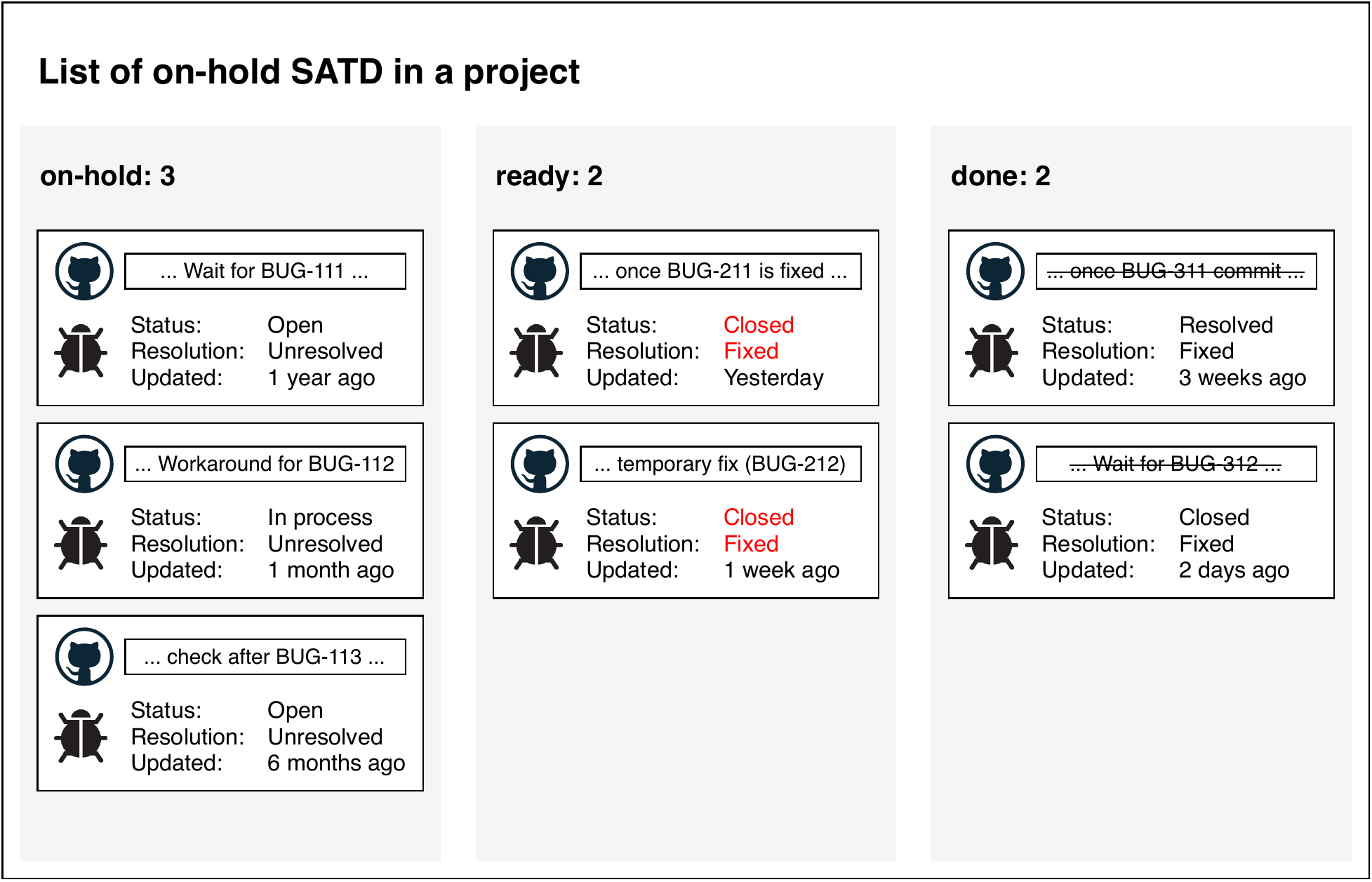}
    \caption{A mockup of On-hold SATD identification tool}
    \label{fig:tool}
\end{figure*}

\subsection{RQ$_3$: To what extent can our approach identify ``ready-to-be-removed'' On-hold SATD?}

To understand how well our approach performs in identifying ``ready-to-be-removed'' On-hold SATD comments, we reported six identified cases to developers in three issue reports, as these six cases correspond to three subsystems of the Apache Hadoop project (two for Hadoop Common, one for Hadoop HDFS, and three for Hadoop YARN). By the time of writing, we have received the feedback from the developers about the two ``ready-to-be-removed'' On-hold SATD comments in the Hadoop Common subsystem.

\tabref{tab:feedback} lists these two instances of On-hold SATD reported to the developers in JIRA issue tracking system %
\footnote{https://issues.apache.org/jira/browse/HADOOP-17047}.

\begin{table}[ht]
    \centering
    \caption{Two ``ready-to-be-removed'' On-hold SATD comments which received developers' feedback}
    \label{tab:feedback}
    \begin{tabular}{lp{7.5cm}}
    \toprule
    \textbf{\#} & \textbf{Ready On-hold SATD} \\
    1 & ``/* return type will change to AFS once HADOOP-6223 is completed */'' \\
    \midrule 
    2 & ``... This should be made deprecated along with the mapred package HADOOP-1230. ... '' \\
    \bottomrule
    \end{tabular}
\end{table}

For the first case, the return type had already changed to AFS, and the resolution of the referred issue ``HADOOP-6223'' had been set to ``resolved''. In the issue report, we suggested that this On-hold SATD comment should be removed. Developers agree that it can be removed: 
\begin{quote}
    \textit{``I think this is correct finding. Would you like to put a patch for this''}
\end{quote}
\edit{Later on, the patch we submitted got integrated into the repository.}

Regarding the second case, we found that the referred issue ``HADOOP-1230'' had also been resolved. Thus, we suggested that developers could apply corresponding changes (\ie making the {\tt setJobConf} method deprecated). The developers agreed that the action should be taken but it is a rather complicated fix, thus recommending a new JIRA issue thread:
\begin{quote}
    \textit{``...we need to update the document in a separate jira.''}\\
    
    \textit{``... Given that is a bigger subject than this fix,  we should discuss on that separately ...''}
\end{quote}

Overall, the two cases for which we have already received feedback on indicate the practical value of our approach for On-hold SATD identification and removal.

\subsection{Replication} \label{appendix}

To facilitate replication, we released our dataset in our online appendix, which can be accessed at \url{https://tinyurl.com/onholdissue}. The spreadsheet file of our dataset contains three sheets: removed comments, \edit{remaining} comments, and identified ``ready-to-be-removed'' On-hold SATD. For all the comments in our dataset, we include the comment context, code file path, line number, referred issue, and our annotation (On-hold SATD or cross-reference). For removed comments, we also include when the code comment was introduced and removed. For the ``ready-to-be-removed'' On-hold SATD, there are also the status and the resolution of the corresponding issues.

\section{Towards a On-Hold SATD Recommender}\label{sec:recommender}

Our findings can serve as guideline for developers writing reference issues in code comments:

\begin{itemize}

    \item Developers should check SATD comments referring to issues which had already been resolved, as we reported that 13\% of comments were removed with a delay of more than one year.
    
    \item When the code comments refer to issues, developers should clearly mention the intention in the comments, \ie whether the issue is used for documentation or to denote the condition on which one is waiting on.

\end{itemize}

While we plan on expanding our work to analyze more projects and to include also other issue tracking systems, we believe that our work can be synthesized into a recommender system for On-Hold SATD. In \figref{fig:tool} we depicted a mock-up of such a recommender. 

The tool would report the list of On-hold SATD comments, ready to be addressed On-hold SATD comments, and removed On-hold SATD. Each item would include comments, links to the original comments and to the pertaining issue (including its status and duration).  

\section{Threats to Validity}

Threats to \emph{construct validity} concern the relation between the theory and the observation, and in this work they are mainly due to the measurements we performed:

\begin{itemize}

\item \emph{Imprecisions in the identification of issue references in comments}. We used the regular expressions in \tabref{regex} to mine issue references in code comments. The regular expressions have been defined and tested by the first author, and are customized for each of the issue trackers used by the subject systems. 

\item \emph{Subjectivity/errors in the manual classification}. To mitigate this threat, the first two authors independently classified the 1,530 issue-referencing comments as On-hold SATD or as cross-reference. Then, the third author resolved the conflicts.

\end{itemize}

Threats to \emph{external validity} concern the generalizability of results. Rather than going large-scale, we preferred to work on a set of ten well-known Java open source projects and to manually validate all issue-referencing comments we found in them in such a way to increase the reliability of the presented data. Other systems should be included in the analysis to allow for a broader generalizability of our conclusions. Also, the results of RQ$_3$ are based on only two feedback we received from developers, thus do not allowing any sort of generalizability but only serving as pointers for qualitative analysis.

\section{Conclusion}

Since the definition of the term ``technical debt'' by Cunningham three decades ago \cite{10.1145/157709.157715}, researchers have investigated the phenomenon, leading to the understanding that it is its creeping, barely visible nature that leads to maintenance and evolvability problems down the road. Developers cannot be faulted for the introduction of technical debt, as software industry functions under great time and budget pressure, and compromises have to be made to meet said time and budget constraints. Indeed, developers often admit that they are creating technical debt, which led to the term ``self-admitted technical debt'' (SATD) coined by Potdar and Shihab \cite{10.1109/ICSME.2014.31}.

A particular type of SATD is the one we named ``On-hold'' SATD, where a developer has to make a compromise or halt development because of an external condition. Human nature dictates that often On-hold SATD is simply forgotten about.

We performed an empirical study to understand whether On-hold SATD can be automatically detected: We analyzed ten open source projects, and found that 8\% of the comments referring to issues are On-hold SATD. To identify On-hold SATD, we developed a classifier using n-gram and auto-sklearn, resulting in an average precision of 0.79, an average recall of 0.70, an average $F_1$-score of 0.73, and an average AUC of 0.97. In short, On-hold SATD can indeed be detected automatically in a fairly reliable way.

To understand how On-hold SATD evolves, we looked into life-span of removed issue-referring comments. We found that the median life-span of On-hold comments is 42 days. This is certainly beyond the horizon of human short-term memory, and indeed we found that after the issues were resolved, 13\% of On-hold SATD takes longer than one year to remove. To evaluate the reliability in identifying On-hold SATD which should be removed, we collected feedback from developers from open source projects. Developers agreed with our findings that the reported On-hold SATD should be fixed or removed.

The next logical step is thus the design and implementation of the recommender system we described in \secref{sec:recommender} and aimed at facilitating the identification, understanding, and resolution of On-hold SATD instances. 

\section*{Acknowledgement}

We gratefully acknowledge the financial support of Japan Society for the Promotion of Science for the JSPS KAKENHI Grant No. 16H05857 and 20H05706, and the Swiss National Science Foundation for the project SENSOR (SNF-JSPS Project No. 183587).

\newpage

\bibliographystyle{IEEEtran}
\bibliography{bibtex}

\end{document}